\documentclass[dvipsnames]{ecai} 

\usepackage{latexsym}
\usepackage{amssymb}
\usepackage{amsmath}
\usepackage{amsthm}
\usepackage{booktabs}
\usepackage{enumitem}
\usepackage{graphicx}

\usepackage{svg}
\usepackage{xcolor}
\usepackage{multirow}
\usepackage{xr-hyper}
\usepackage{hyperref}

\usepackage[doi=false,url=false]{biblatex}
\newbibmacro{string+doi}[1]{%
  \iffieldundef{doi}{\iffieldundef{url}{#1}{\href{\thefield{url}}{#1}}}{\href{http://dx.doi.org/\thefield{doi}}{#1}}}	
\DeclareFieldFormat*{title}{\usebibmacro{string+doi}{\mkbibemph{#1}}}

\newcommand{\BibTeX}{B\kern-.05em{\sc i\kern-.025em b}\kern-.08em\TeX}

\bibliography{lamarl}

\begin{document}


\begin{frontmatter}



\title{Towards Language-Augmented Multi-Agent Deep Reinforcement Learning}


\author[A,B]{\fnms{Maxime}~\snm{Toquebiau}\thanks{Corresponding Author. Email: maxime.toquebiau@sorbonne-universite.fr}}
\author[B]{\fnms{Jae-Yun}~\snm{Jun}}
\author[A]{\fnms{Faïz}~\snm{Benamar}} 
\author[A]{\fnms{Nicolas}~\snm{Bredeche}} 

\address[A]{Sorbonne Universit\'{e}, CNRS, ISIR, F-75005 Paris, France}
\address[B]{ECE Paris, France}


\begin{abstract}
Most prior works on communication in multi-agent reinforcement learning have focused on emergent communication, which often results in inefficient and non-interpretable systems. Inspired by the role of language in natural intelligence, we investigate how grounding agents in a human-defined language can improve the learning and coordination of embodied agents. We propose a framework in which agents are trained not only to act but also to produce and interpret natural language descriptions of their observations. This language-augmented learning serves a dual role: enabling efficient and interpretable communication between agents, and guiding representation learning. We demonstrate that language-augmented agents outperform emergent communication baselines across various tasks. Our analysis reveals that language grounding leads to more informative internal representations, better generalization to new partners, and improved capability for human-agent interaction. These findings demonstrate the effectiveness of integrating structured language into multi-agent learning and open avenues for more interpretable and capable multi-agent systems.\footnotemark 
\end{abstract}

\end{frontmatter}

\footnotetext[1]{Our code is available at \href{https://github.com/MToquebiau/LAMARL}{github.com/MToquebiau/LAMARL}}


\section{Introduction}\label{sec:Intro}

Multi-agent deep reinforcement learning (MADRL) studies how groups of artificial agents can be trained to optimize their performance on a given task.
One crucial skill for facilitating social interactions is communication, as observed in many social animal species~\cite{Smith2003_AnimalSignals, Searcy2010_AnimalCommEvo}, and, of course, humans~\cite{Greene2003_Communication}. It enables a great range of social abilities: from sharing information about local observations to negotiation, knowledge transmission, and teaching. 
Inspired by these natural systems, artificial agents can benefit from learning to communicate, improving coordination and overall group performance.

A central element of human communication is natural language. Through the use of shared sets of symbols and rules, humans are able to converse efficiently about diverse situations with other, possibly unknown, individuals. Thanks to their compositional nature, natural languages enable efficient information transmission, allowing the expression of complex ideas in a compressed form using a rather limited set of different symbols~\cite{Pinker1990_NaturalLang}. 
Additionally, language plays a key role in the intellectual development of children by enabling them to describe their environment, internalize others' experiences, and build upon structured linguistic rules~\cite{Vygotsky1934, Piaget1978_Symbole, Tomasello2009_Cultural}. 
By learning to associate concepts from the world to language, one can better understand the different components of the world, how they can relate, and how to build logical reasoning from them. 

Following these observations, a promising avenue for artificial intelligence is in the investigation of how language can help artificial agents when learning to behave in complex environments and social settings. Many research works have pursued this direction in the context of single-agent reinforcement learning, using language to guide training by describing the world~\cite{Hill2021_Grounded, Carta2023_GLAM}, providing and exploring high-level goals~\cite{Colas2020_Imagine}, and decomposing tasks~\cite{Hu2019_HierarLang}. However, language-augmented training in multi-agent settings is still an understudied subject. The dominating approach in MADRL research is emergent communication, where agents develop their own communication system from scratch in the course of training~\cite{Zhu2024_MACSurvey}. While this can be effective, it often leads to inefficient communication systems~\cite{Chaabouni2019_AntiEfficient, Galke2022_Emergent} and lacks the guidance provided by a pre-defined set of symbols and rules to describe the world and share information with anyone. 
Some works have studied learning to communicate with natural language~\cite{Lazaridou2020_MACNatLang, Gupta2021_DynamicPop}, but very few of them do so in embodied settings where agents explore their environment through physical interactions. 
Finally, language is also at the center of the recent trend focused on using large language models (LLMs) in various social settings~\cite{Bakhtin2022_CICERO, Park2023_LLMtown}. While this direction offers many promising results, it focuses on using pre-trained language models, thus missing the insights on how language learning and behavior learning interact.

In this paper, we make a step towards filling this gap by studying language-augmented agents trained to solve multi-agent tasks with MADRL. 
We argue that \textbf{learning to use a human-defined language is beneficial for developing intelligent agents capable of complex reasoning in embodied and social settings}. To back this claim, we propose a method to teach a pre-defined language to MADRL agents, by training them to produce language descriptions of their observations. In this framework, agents concurrently learn to solve a multi-agent embodied task and to use and understand the given language. Thus, language training serves a dual purpose: (i) \textbf{learning language-based communication}, while (ii) \textbf{guiding representation learning} by providing an efficient way of extracting valuable information from the world. To allow this, we provide agents with language examples that show how the pre-defined language should be used to describe observations. This data is used to train supervised language objectives concurrently to training on the multi-agent reinforcement learning objective.

Through extensive experiments, we study the many advantages provided by this language-augmented training scheme. Language-augmented agents are shown to outperform emergent communication baselines across multiple tasks. We provide various results indicating that language training helps agents build better representation, including embedding visualizations and ablation studies. Further experiments showcase multiple benefits of language-augmented multi-agent systems for generalizing to new partners and interacting with humans. Finally, we provide a discussion on the costs, benefits and various potentials of this approach, and argue for more works in this direction.

\section{Related Works}\label{sec:RelatedWorks}
 
\paragraph{Emergent communication.}
The main trend for learning communication with MADRL is emergent communication, in which agents develop their own communication protocol during training. 
While emergent communication has been used for a few decades as a model to explore the origins and evolution of languages~\cite{Steels1997}, recent MADRL works have employed it in a more prescriptive manner to enable end-to-end learning of communication from reinforcement learning signals~\cite{Zhu2024_MACSurvey}. With messages being generated and processed only by differentiable neural networks, emergent communication can be learned with deep reinforcement learning as a way to maximize the returns of all communicating agents~\cite{Foerster2016_DIAL,Sukhbaatar2016_CommNet}. A large body of work have proposed methods to improve emergent communication by allowing targeted communication~\cite{Das2019_TarMAC, Singh2019_IC3Net}, limiting bandwidth~\cite{Kim2018_SchedNet, Han2023_MBC}, promoting social influence~\cite{Jaques2019_SocialInfluence}, or improving learning stability~\cite{Zhang2019_VBC}. While these methods are shown to improve task performance, multiple works have highlighted their limitations with regards to efficiency of communication~\cite{Chaabouni2019_AntiEfficient, Galke2022_Emergent}, measuring communicative performance~\cite{Lowe2019_Pitfalls}, and interpretability~\cite{Bouchacourt2018_HowAgentsSee}.

\paragraph{Grounding communication.}
To address these issues, one approach is to ground the emergent communication systems in external modalities used as a source of meaning. 
To ensure that messages carry information about the environment, agents can be trained to maximize the informational content of their messages by reconstructing input observations~\cite{Lowe2020_S2P, Lin2021_GroundMAC, Karten2023_CompoConcept}. 
Natural language can also be used as a source of meaning. By learning to associate observations with language descriptions, agents learn to extract key features of their input space to communicate more effectively~\cite{Havrylov2017_EmergenceLang, Lazaridou2020_MACNatLang, Tucker2021_DiscreteEC, Li2024_LangGround}. These grounding mechanisms complement reinforcement learning by introducing auxiliary supervised objectives that guide agents toward learning meaningful communication. However, they still produce emergent languages that may carry useless information, are specialized to their training setting, and are difficult to interpret. 

\paragraph{Multi-agent LLMs.}
The recent progress in large language models (LLMs) has produced models with strong linguistic and human-like reasoning capabilities. This has led to a growing body of work on building collectives of LLM-based agents for solving multi-agent tasks~\cite{Park2023_LLMtown, Guo2024_LLMsMA}. By prompting the LLMs in particular ways, these agents can be assigned distinct personas, goals, or roles, and can be specialized for particular sub-tasks within a larger coordination problem~\cite{Perez2024_CulturalEvo}. A particularly relevant example is CICERO~\cite{Bakhtin2022_CICERO}, which fine-tunes LLMs with multi-agent reinforcement learning to enable strategic dialogue in a multiplayer strategy game. However, CICERO and other similar approaches rely heavily on the pretrained language capabilities of LLMs to produce rich, high-level interactions. Therefore, they decouple language generation training from the learning of physical and social skills.
This separation contrasts with research in cognitive and developmental psychology, which emphasizes the intricate relation between linguistic, perceptual, motor, and social development~\cite{Piaget1978_Symbole}. 

\paragraph{Language-augmented reinforcement learning.}
On the other hand, the relationship between learning language and motor skills has been explored in single-agent RL. Researchers have used language as a means to ground perception~\cite{Hanjie2021_EMMA, Hill2021_Grounded, Carta2023_GLAM}, express goals~\cite{Colas2020_Imagine}, and predict future outcomes~\cite{Huang2022_InnerMonol, Lin2023_Dynalang}. Language also supports hierarchical reinforcement learning by enabling high-level policies to generate interpretable sub-goals~\cite{Hu2019_HierarLang, Weir2023_HLLP}. As these advances are mostly confined to single-agent RL, we argue that language can play an equally transformative role in multi-agent settings by supporting structured reasoning, coordination, and generalization in complex interactions.

\begin{figure*}
    \centering
    \includesvg[width=0.95\linewidth]{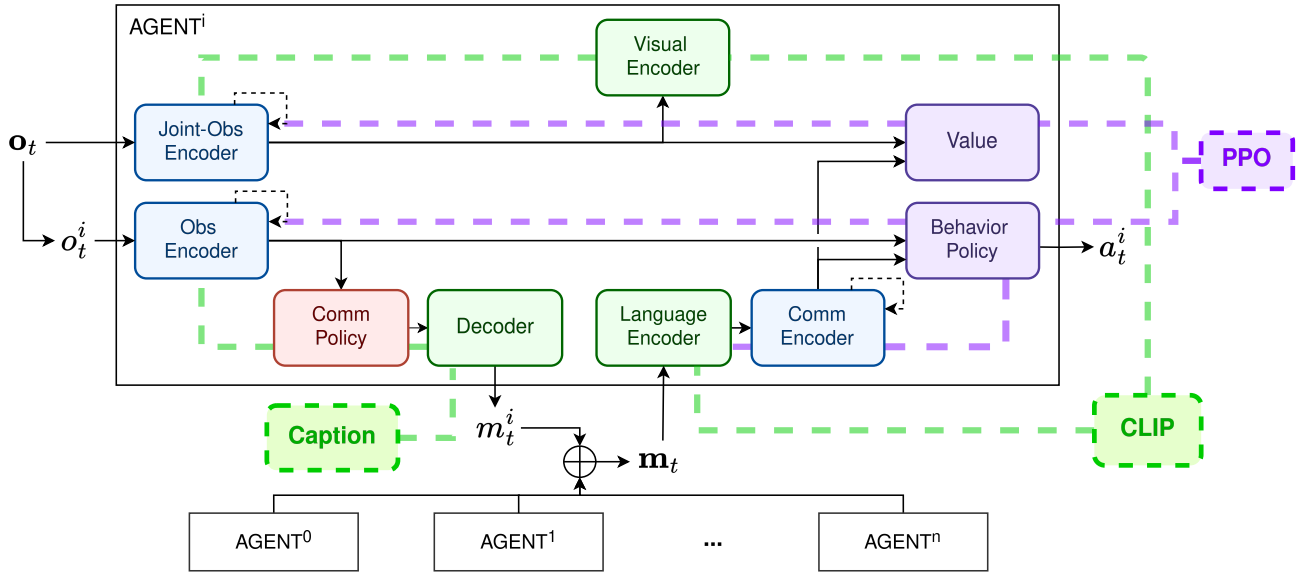}
    \caption{Language-augmented agent architecture. Each module represents a neural network with a specific purpose. \textcolor{RoyalBlue}{\textbf{Encoder modules}} receive incoming information (observations or communication) and embed it in latent representations used in further modules, with dashed self-pointing arrows indicating the use of recurrent neural networks to allow some memorization over multiple time steps. The \textcolor{BrickRed}{\textbf{communication policy}} selects what information in the observation should be communicated. \textcolor{ForestGreen}{\textbf{Language modules}} provide language capabilities: the decoder generates messages, the language encoder transforms incoming messages in a compact latent representation, and the visual encoder is used for training the language encoder (as described in Section~\ref{sec:Learning}). Finally, \textcolor{Plum}{\textbf{reinforcement learning modules}} take information from both observations and communication to select actions and generate values. Colored dashed lines indicate the three training objectives and which modules they impact.}
    \label{fig:archi}
\end{figure*}

\section{Preliminaries}\label{sec:Background}

\paragraph{DecPOMDP with communication.}
We use the decentralized partially-observable Markov decision process (Dec-POMDP) \cite{Oliehoek2016_DecPOMDP} as a formal framework to study multi-agent cooperative settings. It is defined as a tuple $\langle\mathbf{S},\mathbf{A},T,\mathbf{O},O,R,n,\gamma\rangle$ with $n$ the number of agents. $\mathbf{S}$ is the set of global states $s$ of the environment. $\mathbf{O}$ is the set of joint observations, with one joint observation being comprised of one local observations $o^i$ per agent: $\mathbf{o}=\{o^1,...,o^n\}\in\mathbf{O}$. Similarly, $\mathbf{A}$ is the set of joint actions, with one joint action $\mathbf{a}=\{a^1,...,a^n\}\in\mathbf{A}$. $T$ is the transition function defining the probability $P(s'|s,\mathbf{a})$ to transition from state $s$ to next state $s'$ with the joint action $\mathbf{a}$. $O$ is the observation function defining the probability $P(\mathbf{o}|\mathbf{a},s')$ to observe the joint observation $\mathbf{o}$ after taking joint action $\mathbf{a}$ and ending up in $s'$. $R:\mathbf{O}\times\mathbf{A}\rightarrow\mathbb{R}$ is the reward function defining the shared reward that agents get at each time step. Finally, the discount factor $\gamma\in[0,1)$ controls the importance of immediate rewards against future gains. On top of this framework, we define communication as taking place during the action-selection process. At each time step, agents receive local observations and generate a message using a learned function: $m^i_t=f^i(o^i_t)$. Messages are then shared through a broadcasting communication channel: each agent receives the concatenation of all produced messages, $\mathbf{m}_t=\{m^1_t,...,m^n_t\}$, and then selects its action. 

\paragraph{Centralized values and decentralized policies.}
To learn multi-agent behaviors, we use algorithms in the centralized training with decentralized execution (CTDE) paradigm. During training, agents are allowed to use some centralized information to improve the learning performance and stability. However, during execution, agents are completely distributed, meaning they choose actions based only on their local observations. In our case, following previous state-of-the-art methods~\cite{Lowe2017_MADDPG, Yu2021_MAPPO}, agents learn decentralized policies conditioned on local observations and incoming messages: $a^i_t\sim\pi^i(o^i_t, \mathbf{m}_t)$; and centralized value functions conditioned on joint observations and incoming messages: $V^i(\mathbf{o}_t, \mathbf{m}_t)\in\mathbb{R}$. The policy and value functions are parameterized by deep neural networks, each agent having its own set of parameters $\theta^i$.  

\section{Language-Augmented Agents}\label{sec:Method}


\subsection{Motivation}
Following finding from developmental psychology, we propose to allow agents to be guided by language during training. Natural languages support cognitive development in various ways. By using symbols to describe concepts from the world, they allow humans to categorize objects from an early age~\cite{Ferry2010_Categorization}. 
By arranging these symbols with rules to combine and compose these concepts, they enable complex reasoning, information transmission, and great generalization abilities~\cite{Piaget1978_Symbole}. Previous works have demonstrated that these advantages of language do not come as simple byproduct of cognition, but rather plays an important role in its development~\cite{Lupyan2012_WhatWordsDo}. 

To replicate this in artificial learning agents, we need a method for combining the learning of language and behavioral skills. For neural networks-based agents, this means concurrently optimizing language and reinforcement learning objectives and having the parameters trained with gradients from both. This way, the internal representations learned inside the neural networks will be shaped by this combined training scheme. In this Section, we present our approach for building a compact language-augmented agent architecture that combines MADRL and language processing techniques. In Section~\ref{sec:Expes}, we will present experimental findings demonstrating the benefits of this approach for improving training and communication in MADRL. 

\subsection{Agent Architecture}\label{sec:Archi}

To design language-augmented agents, we build a neural network architecture, shown in Figure~\ref{fig:archi}, that integrates the tools required for learning to (i) complete a given embodied task and (ii) use and understand a given language. Problem (i) is addressed using classical elements of MADRL. Each agent $i$ is equipped with a \textit{local behavior policy} $\pi^i(o^i_t)$, and a \textit{centralized value function} $V^i(\mathbf{o_t})$, as described in Section~\ref{sec:Background}. Prior to computing the policy and value, the local and joint observations are passed through \textit{encoders} -- "Obs Encoder" and "Joint-Obs Encoder", respectively -- to extract valuable information. The two observation encoders are defined as recurrent neural networks to allow some memorization of information gathered in previous time steps. With only these four modules, an agent can learn to behave in a multi-agent setting, but has no communication capacities. 

To address problem (ii), that is, enabling the learning of language-based communication, agents are equipped with additional neural network modules. To generate messages based on the information contained in local observations, the \textit{communication policy} first extracts valuable information to convey from the encoded local observation. Its output is passed to the \textit{decoder} that generates the message using the symbols and rules of the learned language. Messages from all agents are concatenated into a single broadcast message $\mathbf{m}_t$ that is embedded to a latent vector representation by the \textit{language encoder}. Finally, the \textit{communication encoder} takes the incoming information and encodes it again to extract the information required by the policy and value functions. The communication encoder is also defined as a recurrent neural network to keep some information exchanged in previous time steps. 


\subsection{Policy and Language Learning}\label{sec:Learning}

The proposed architecture is learned by optimizing multiple objectives concurrently. The behavior policy and value function are learned following the multi-agent proximal policy optimization (MAPPO) algorithm~\cite{Yu2021_MAPPO}. Language modules are trained with two distinct supervised learning objectives, captioning~\cite{Herdade2019_Captioning} and contrastive learning~\cite{Radford2021_CLIP}, which are both commonly used in the natural language processing domain. 

\paragraph{Multi-agent PPO.}
Concretely, the reinforcement learning objective defined by MAPPO is two-fold. For any agent $i$, the policy $\pi^i$ is trained to maximize the clipped surrogate objective defined by PPO~\cite{Schulman2017_PPO}, denoted as $\mathcal{L}^\pi(\theta^i)$. The value function $V^i$ is trained to minimize a clipped mean-squared Bellman error loss, noted $\mathcal{L}^V(\theta^i)$ (details on both losses are given in Appendix A). Intuitively, the policy learns to produce actions that maximize global returns, while the value learns to estimate these returns. 

\paragraph{Generating language with captioning.} 
To learn language-based communication, agents have to learn how to describe their observations with language utterances. This resembles the classical natural language processing task of image captioning, where a language model learns to describe scenes presented in input images. In our case, the agents are provided with language descriptions of their local observations. Given observation--descriptions pairs $(o, l)$, with $l=(t_0,...,t_K)$ a sequence of tokens, the decoder learns to generate $l$ based on the information given by the communication policy, by minimizing the cross-entropy loss:
\begin{equation}
    \mathcal{L}^{capt}(\theta^i)=-\sum_{k=1}^Kp(t_k)\log\hat{p}(t_k|o,t_0,...,t_{k-1},\theta^i),
    \label{eq:Captioning}
\end{equation}
with $p$ the target probability density and $\hat{p}(\cdot|o,\theta^i)$ the probability density generated by the decoder. 

\paragraph{Encoding language with contrastive learning.}
To learn to understand incoming language messages, agents need to learn to transform sequences of tokens into latent vector representations. To this end, the language encoder $E^L$ and the visual encoder $E^V$ are trained jointly on the contrastive learning objective defined by CLIP (contrastive language-image pre-training)~\cite{Radford2021_CLIP}. Given a batch of joint-observations--descriptions pairs $(\mathbf{o}, \mathbf{l})$, the encoders are trained to minimize the following loss:
\begin{gather}
    \mathcal{L}^{CLIP}(\theta^i)=\sum_{j,k}CLIP(j,k),\\
    CLIP(j,k)=
    \begin{cases}
        -cosim(E^V(\mathbf{o}),E^L(\mathbf{l})) & \text{ if }j=k, \\
        cosim(E^V(\mathbf{o}),E^L(\mathbf{l})) & \text{ if }j\neq k.
    \end{cases}
    \label{eq:CLIP}
\end{gather}
The idea behind CLIP is that the two encoders should generate similar latent representations for associated pairs $(\mathbf{o}_j, \mathbf{l}_j)$, and different latent representation for unassociated pairs $(\mathbf{o}_j, \mathbf{l}_k)$. Thus, $\mathcal{L}^{CLIP}(\theta^i)$ simultaneously minimizes the cosine similarity between related representations, and maximizes the unrelated ones. 

\paragraph{Complete objective.}
The complete learning objective optimized by our agents is the following:
\begin{equation}
    \begin{aligned}
        \mathcal{L}(\theta^i) &= \beta^\pi \mathcal{L}^\pi(\theta^i) + \beta^V \mathcal{L}^V(\theta^i) + \beta^{capt} \mathcal{L}^{capt}(\theta^i)\\
        &\quad  + \beta^{CLIP} \mathcal{L}^{CLIP}(\theta^i).
    \end{aligned}
    \label{eq:Loss}
\end{equation}
Because this loss combines multiple objectives with different amplitudes and learning rates, we adopt a dynamic weighting approach, with one weighting parameter $\beta$ per loss, to ensures that all losses are optimized at the same rate (see Appendix B for more details). 

By integrating these language learning objectives into the multi-agent reinforcement learning algorithm, we enable end-to-end training of both policy learning and language skills. Both observation encoders receive gradients from the language losses (as shown in Figure~\ref{fig:archi}). This favors the policy and value functions by encouraging agents to extract meaningful concepts from their observations.

\subsection{Language and Oracle Definition}\label{sec:LanguageDef}

\begin{figure}
    \centering
    \includesvg[width=0.95\linewidth]{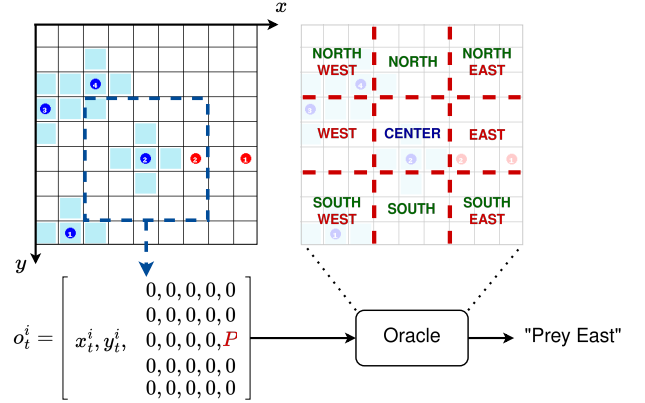}
    \caption{Illustration of the oracle's process for describing observations. In the top-left part is a screenshot of the MA-gym Predator-Prey task (agents in blue and preys in red). Agents observe their position in the grid and the objects in their $5\times 5$ observation range (shown in dark blue). The oracle generates a language description describing the location of observed preys.}
    \label{fig:oracle}
\end{figure}

To enable agents to communicate about their observations, we define a simple language that allows describing elements from the environments. Importantly, the language should only describe elements that are important for the task at hand. For example, in a Predator-Prey setting, agents need to find preys and coordinate to catch them. As agents have a limited field of view, sharing the positions of observed preys to the group should improve the prey capture efficiency. Therefore, we design a simple language that allows this by stating the cardinal position (i.e., "North", "South", "East", "West", and "Center") of the observed preys, as shown in Figure~\ref{fig:oracle}.

To generate language examples, an \textit{oracle} function is built, based on a set of rules, to transform any observations into the corresponding language description. The language can be adapted to fit the communication needs of each task: e.g., with other entities, colors (see Appendix D for more details).


Despite its apparent simplicity, this language features similar properties as more complex natural language: it is structured, combinatorial and compositional. Thus, it provides the tools for efficiently signaling of valuable information. 

\section{Baselines Definition}\label{sec:BaselineDef}

In the experiments presented in the next section, we investigate the benefits of learning to communicate with a pre-defined language by comparing our language-augmented agents, termed \texttt{LAMARL}, with three emergent communication baselines, defined hereafter.

\texttt{EC} agents, for "Emergent-Communication", use a basic differentiable emergent communication strategy. They use the same architecture, as illustrated in Figure~\ref{fig:archi} but with no language modules. Agents use the vector generated by the communication policy as continuous messages to send to their partners. 

\texttt{EC-AutoEncoder} agents build upon the \texttt{EC} agents by grounding their communication system in observation. Inspired by~\cite{Lin2021_GroundMAC}, an auto-encoding objective is employed to ensure that messages carry some information about the input observations. 

\texttt{EC-LangGround} agents implement another grounding method for emergent communication. Inspired by the recent work of~\cite{Li2024_LangGround}, emergent communication is grounded in language by minimizing the distance between the generated continuous messages and the embedding of language descriptions of the input observation. A language encoder is pre-trained beforehand and used during training only to provide target embeddings to compute the grounding objective.

Finally, \texttt{No Comm} agents do not communicate at all, acting as a baseline to measure the impact of communication in the different settings. 



\begin{figure*}[t]
    \centering
     \includesvg[width=0.95\linewidth]{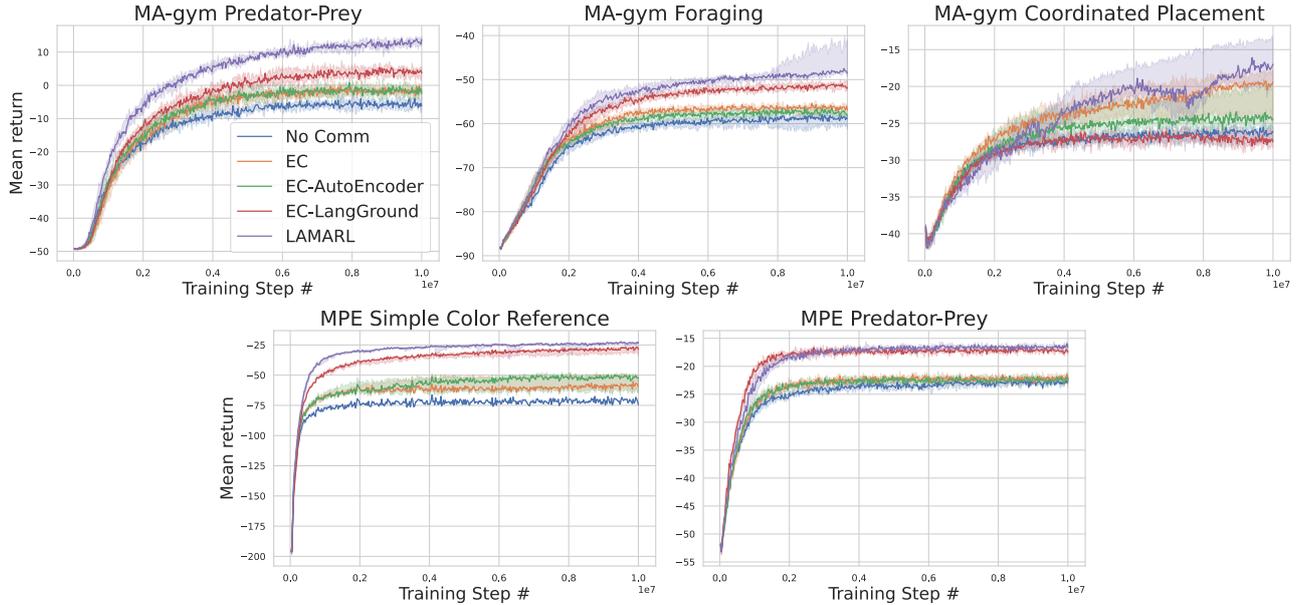}
    \caption{Training performance of LAMARL agents against baselines (7 runs each, with median and 95\% confidence interval). \texttt{LAMARL} agents consistently beat the emergent communication baselines. }
    \label{fig:Train}
\end{figure*}

\section{Experiments}\label{sec:Expes}

To study the advantages provided by our language-augmented architecture, we present a set of experiments that compare it to emergent communication baselines. We first look at training performance in classical multi-agent problems and then provide a deeper analysis of the impacts of language learning on our agents. Important hyperparameters and our code for running all experiments are provided in the Supplementary Materials. 

\subsection{Learning to Communicate}
\label{sec:XP_Training}

We evaluate our approach in the MA-gym environment~\cite{Koul2019_magym} and the Multi-agent Particle Environment (MPE)~\cite{Lowe2017_MADDPG}. MA-gym offers a two-dimensional grid-world setting, while MPE is a continuous environment with continuous states and actions. 
We study five separate tasks:
\begin{itemize}
    \item \textit{MA-gym Predator-Prey}, a discrete predator-prey setting with four agents having to catch two preys. 
    \item \textit{MA-gym Foraging}, a discrete level-based foraging setting where four agents have to coordinate to retrieve colored resources. The color of the resource (yellow, green or purple) indicates the number of agents required to be foraged (one, two or three).
    \item \textit{MA-gym Coordinated Placement}, a discrete cooperative navigation problem where two agents have to find two landmarks of the same color and navigate on them to complete the episode.
    \item \textit{MPE Simple Color Reference}, a colored extension of the "simple reference" task in MPE, where two agents need to navigate onto landmarks, but only the other agent has the information of which landmark to go onto. Thus, communication is required for agents to solve the task.
    \item \textit{MPE Predator-Prey}, a continuous predator-prey setting with two agents having to catch a prey.
\end{itemize}
Details on these tasks are provided in Appendix D. All settings are partially observable, with agents only able to see nearby entities. The MPE Simple Color Reference setting requires communication to be solved, but all of them can be significantly improved by sharing information observed locally.

\paragraph{Results.} The results in Fig.~\ref{fig:Train} show that \texttt{LAMARL} consistently outperforms all baselines across all tasks, demonstrating the advantages of structured language-based communication. \texttt{LAMARL} learns faster in three out of the five tasks, and always stabilizes at a superior performance level than competitors. Across all tasks, agents without communication perform the worst (the \texttt{No Comm} baseline), confirming the benefits of using communication. \texttt{EC-LangGround} dominates other EC variants, but stays far from the performance of \texttt{LAMARL}. In the Coordinated Placement task, where coordination requires precise information sharing, \texttt{EC-LangGround} collapses and provides results comparable to the \texttt{No Comm} baseline, while vanilla \texttt{EC} is nearly competitive with \texttt{LAMARL}. For this task, \texttt{LAMARL} appears to be struggling more, especially in the early training steps, but still succeeds in dominating other approaches in the longer run. Note that this performance is achieved by a significantly smaller bandwidth, with \texttt{LAMARL} sending on average $1.19$ tokens (with one token coded on one integer), compared to emergent communication baselines sending $2$ floating numbers (for \texttt{EC} and \texttt{EC-AutoEncoder}) or $4$ floating numbers (for \texttt{EC-LangGround}).

\begin{figure*}[t]
    \centering
     \includegraphics[width=0.95\linewidth]{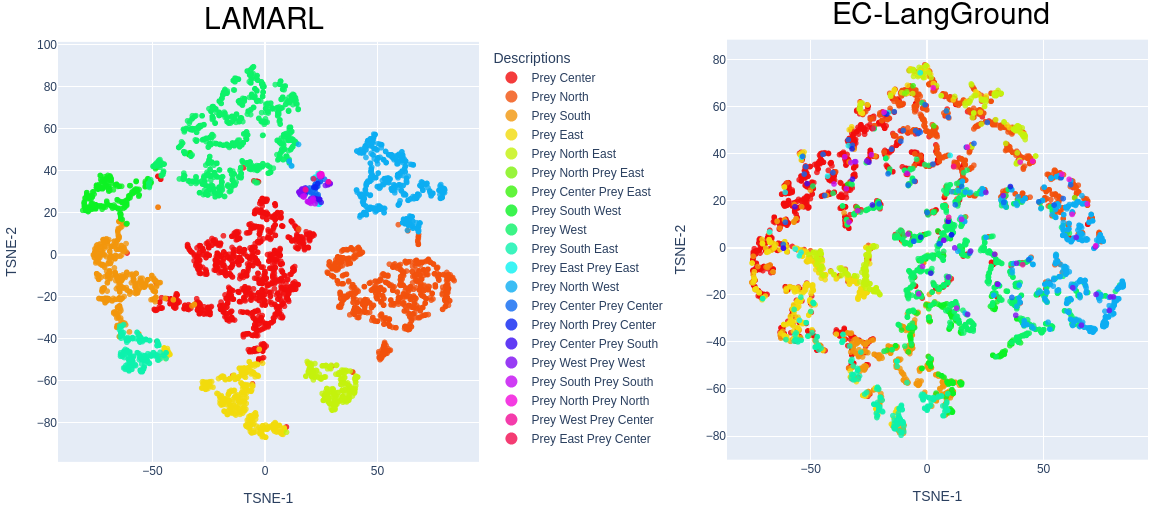}
    \caption{Visualizations of embeddings produced by the communication policy in \texttt{LAMARL} (left) and \texttt{EC-LangGround} (right). By explicitly learning to generate language utterances, \texttt{LAMARL} learns better structured representation, as shown by the more distinct clusters and the silhouette scores: $0.59$ for \texttt{LAMARL} and $0.06$ for \texttt{EC-LangGround}.}
    \label{fig:embeds}
\end{figure*}

\subsection{Internal representations}
\label{sec:Embeds}

To study how learning language during training affects the internal representations of our agents, we study the learned embedding spaces. We retrieve a set of observations from the MA-gym Predator-Prey task and the corresponding descriptions for labeling. We encode these observations with the learned weights of evaluated agents. To plot the generated embeddings, we use the t-SNE dimensionality reduction technique~\cite{Vandermaaten2008_tSNE} and plot the encoded points with a color referring to the corresponding description. Figure~\ref{fig:embeds} shows the result of this procedure, comparing \texttt{LAMARL} to the best emergent communication baseline \texttt{EC-LangGround}. The embeddings present in this Figure were generated by the communication policy, meaning that they are directly involved in communication: by being the input of the decoder for \texttt{LAMARL} (of dimension 16) and the message itself for \texttt{EC-LangGround} (of dimension 4).

\paragraph{Results.}
The visualizations in Figure~\ref{fig:embeds} reveal that \texttt{LAMARL} produces significantly more structured embedding spaces than \texttt{EC-LangGround}, illustrating the stronger inductive bias introduced by language supervision. In particular, \texttt{LAMARL} consistently exhibits well-separated clusters corresponding to distinct environmental conditions, with cardinal directions (i.e., "North", "South", "East", "West") clearly distributed around a central point (i.e., "Center"), mirroring the spatial layout of the task. Note that this particular disposition is not a product of chance as it was found in all evaluated runs. This demonstrates the interaction between language and policy training: observations close to each other in space will be represented close in the embedding space to highlight their relationship. 
Additionally, the embedding space reflects the compositional structure of language, as intercardinal directions (e.g., "North East", "South West") appear positioned between their corresponding cardinal clusters. By contrast, the embeddings from \texttt{EC-LangGround} are notably less organized. While there is some grouping of similar locations, the clusters are more diffuse and lack the spatially coherent arrangement observed in \texttt{LAMARL}. This contrast suggests that the emergent language in \texttt{EC-LangGround} is not subject to the same structuring pressure on internal representations as the grounded linguistic supervision in \texttt{LAMARL}. In Appendix E, we provide similar visualizations for the embeddings generated inside the observation encoder, demonstrating that this structuring effect impacts the agents' network even more profoundly.


\begin{table}[h]
    \centering
    \begin{tabular}{ccc}
    \hline
    Agent version & \begin{tabular}[c]{@{}c@{}}Language\\ learning\end{tabular} & \begin{tabular}[c]{@{}c@{}}Communication\\ strategy\end{tabular} \\ \hline
    LAMARL & Yes & Learned language  \\
    Lang+Oracle & Yes & Oracle  \\
    Lang+No Comm & Yes & No communication  \\
    No Lang+Oracle& No & Oracle  \\ 
    Observations& No & Raw observations  \\
    No Comm & No & No communication  \\ \hline
    \end{tabular}
    \vspace*{1ex}
    \caption{Properties of variants compared in the ablation study.}
    \label{tab:Ablat}
\end{table}

\subsection{Ablation Study}
\label{sec:XP_Ablat}

\begin{figure}
    \centering
     \includesvg[width=0.95\linewidth]{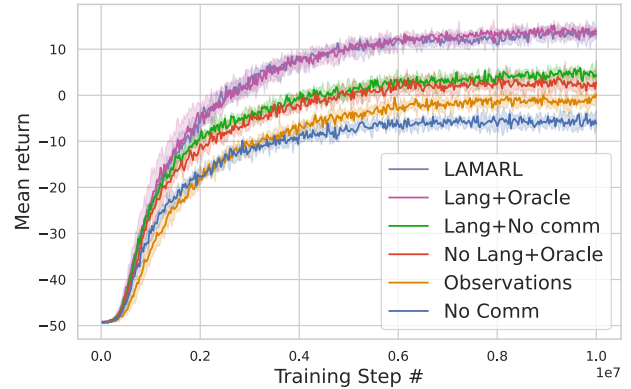} 
    \caption{Training performance of the ablated version on Predator-Prey $18\times 18$ (7 runs each, with median and 95\% confidence interval). \texttt{LAMARL} agents achieve similar performance as \texttt{Oracle} agents, showing that they succeed in using the language properly. All other ablations are inferior.}
    \label{fig:Ablat}
\end{figure}

To demonstrate how learning language helps agents beyond just providing an efficient communication system, we conduct an ablation study over two main properties: (i) whether agents are trained on the language objectives or not, and (ii), how agents communicate.
These properties are summarized in Table~\ref{tab:Ablat}. \texttt{Lang+Oracle} learns language, but sends the "perfect" messages given by the oracle instead of the messages generated by the agents. \texttt{Lang+No comm} learns language but do not communicate, while \texttt{No Lang+Oracle} does not learn language but sends the oracle's "perfect" messages. 
Finally, \texttt{Observations} does not learn language and sends raw observations as messages. 

\paragraph{Results.}
Results of this ablation study, shown in Figure~\ref{fig:Ablat}, offer many insights on how learning language impacts training. First, \texttt{LAMARL} and \texttt{Lang+Oracle} achieve similar performance across training, showing that our agents properly learn to use language. Second, and perhaps most importantly, \texttt{Lang+No Comm} and \texttt{Lang+Oracle} are both better than their variant that do not learn language, i.e., \texttt{No Comm} and \texttt{No Lang+Oracle}, respectively. These two results collectively demonstrate that, regardless of the communication strategy (here either none or perfect), being grounded in language during training improves the task performance. Finally, the \texttt{Observations} baseline performs worse than all other variants using language-based communication. This indicates that, while the informational content in the raw observations is maximal, the lack of compression makes the learning of information transmission more difficult. This corroborates previous results that highlighted how compression and compositionality were crucial properties of languages, enabling to be learned more easily~\cite{Kirby2015_CompressionComm, Chaabouni2020_Composit}.
\subsection{Zero-Shot Teaming}
\label{sec:XP_ZST}

\begin{figure}[t]
    \centering
     \includesvg[width=\linewidth]{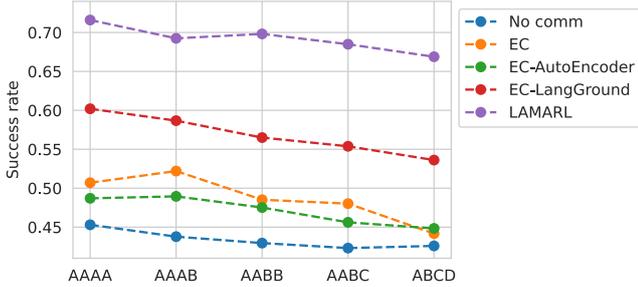}
    \caption{Evaluation performance in the zero-shot teaming experiment. Letters indicate the team composition. Results show the success rate over the evaluation run ($1.2\times10^6$ episodes). 
    }
    \label{fig:ZST}
\end{figure}

Next, we evaluate whether language-based communication allows better cooperation with previously unknown agents. For each method, we compose teams of four agents of various degrees of heterogeneity, ranging from a highly homogeneous team where all agents were trained together to a highly heterogeneous team where agents are picked from different runs with the same algorithm (e.g. one agent from each of the four best teams). As teams are composed of four agents, we devise five possible arrangements termed AAAA (fully homogeneous), AAAB, AABB, AABC, ABCD (fully mixed) and test these configurations for each variant. We use the original Predator-Prey task ($18\times18$). Each team composition is evaluated over $25,000$ episodes, recording the success rate (all variants are evaluated on the same initial conditions). 
We identify two possible reasons why zero-shot teaming degrades performance: (1) a mismatch in communication norms: agents may no longer understand each other due to differences in learned communication protocols, and (2) a mismatch in coordination norms: coordination strategies may fail to transfer effectively between mixed teams.

\paragraph{Results.} The results of the zero-shot teaming experiment in Fig.~\ref{fig:ZST} confirm that mixing agents from different trained teams generally degrades performance, except for the \texttt{No Comm} baseline, which is largely unaffected. The \texttt{No Comm} agents do not rely on communication, yet still experience a slight decline in performance, possibly due to a mismatch during coordination strategies. The \texttt{EC} and \texttt{EC-AutoEncoder} agents exhibit a clear performance drop as heterogeneity increases, likely a result of both a loss of communication and coordination norms, though it is difficult to separate their contributions. \texttt{EC-LangGround}, which grounds communication in language, suffers a similar loss to \texttt{EC}, suggesting that the structured grounding does not provide enough stability to maintain cross-team compatibility. \texttt{LAMARL} remains the strongest overall, showing the best success rate across all team compositions, and only a moderate decrease in the face of heterogeneity. 

%

\subsection{Interaction}
\label{sec:XP_Interact}

\begin{figure}[t]
    \centering
     \includesvg[width=0.9\linewidth]{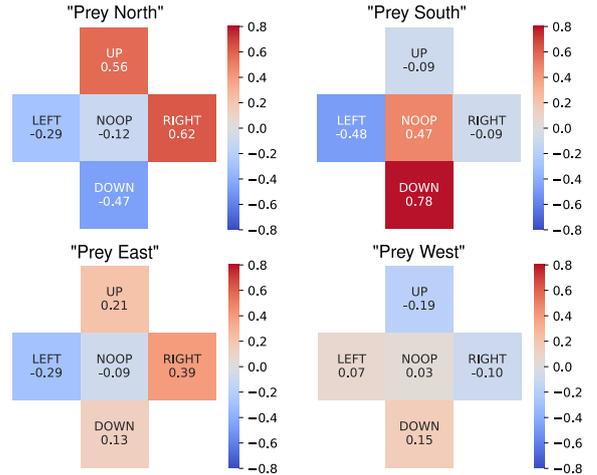}
    \caption{Change in action probabilities depending on the input message. Positive values (resp., negative values) indicate that the action will be taken more often (resp., less often) if the corresponding message is received, compared to if no message is received. } 
    \label{fig:Interact_ActProbs}
\end{figure}

Finally, we present an interaction experiment to demonstrate the interaction capabilities of our language-augmented agents. Because \texttt{LAMARL} agents are trained to use a human-defined language for communication, human-agent interaction should be rather straightforward. To demonstrate this, we manually injected directional messages (i.e.,~‘north’, ‘south’, ‘east’ or ‘west’) to evaluate the agents’ reactions. Agents from a team trained on Predator-Prey are put in an empty setting and given locations for a supposedly observed prey. In Figure~\ref{fig:Interact_ActProbs}, we plot the change in action probability relative to when agents do not receive any message. To compute it, we normalize the action probabilities with message by the action probabilities without message and center the result around zero (see Appendix F for raw action probabilities with and without messages). Thus, positive values correspond to an increase in action probabilities and, conversely, negative values to a decrease. 

\paragraph{Results.} Figure~\ref{fig:Interact_ActProbs} shows the positive results of this interaction experiment: agents react accordingly to the inserted messages; e.g., if a prey is announced North, agents will tend to go towards this location more often than if no message is received. The impact is less clear in some cases (e.g., "Prey West"), which is explained by the fact that these particular agents have a learned bias toward choosing the "LEFT" action more often when no message is received (as shown in Appendix F). These results demonstrate that the language taught to agents can successfully impact their behavior.

\section{Discussions}\label{sec:Discussions}

We introduced \texttt{LAMARL}, a language-augmented multi-agent deep reinforcement learning framework that extends a standard MADRL architecture with minimal yet effective components for learning to understand and generate language. By grounding the agents' internal representations in linguistic input, our method enables them to develop more structured and semantically meaningful embeddings of their environment. This grounding leads to faster learning, improved task performance, and more efficient, interpretable communication among agents.

Our comparison with the \texttt{EC-LangGround} baseline highlights a key distinction: whereas \texttt{EC-LangGround} relies on mimicking a pretrained language encoder to produce messages, \texttt{LAMARL} jointly trains language understanding, generation, and behavior. This results in agents that not only acquire communicative competence but also benefit from deeper integration of language into their reasoning processes. 

This work demonstrates some of the benefits of incorporating language learning into embodied multi-agent systems. While language is a central element in contemporary AI, its role as a grounding signal during training remains underexplored in multi-agent settings -- despite its importance in natural social interactions. \texttt{LAMARL} takes a first step toward bridging this gap, and we hope it inspires future research at the intersection of language, embodiment, and cooperation. Promising directions include scaling to more complex environments, exploring different communication protocols, and integrating richer forms of natural language interaction.





\begin{ack}
The authors appreciate the ECE for financing the Lambda Quad Max Deep Learning server, which is employed to obtain the results illustrated in the present work. 
\end{ack}



\printbibliography

\onecolumn

\appendix
\addcontentsline{toc}{section}{Appendices}
\section*{Appendices}

\section{Proximal Policy Optimization Loss}\label{app:PPO}

Here we provide details on the training objectives for the Proximal Policy Optimization (PPO) algorithm~\cite{Schulman2017_PPO}. PPO is a policy gradient method that improves stability by limiting the size of policy updates. The core idea is to maximize a clipped surrogate objective that discourages large policy deviations. Let $\pi_\theta$ denote the current policy parameterized by $\theta$, and $\pi_{\theta_{\text{old}}}$ the policy before the update. The probability ratio is defined as:
\[
r_t(\theta) = \frac{\pi_\theta(a_t | s_t)}{\pi_{\theta_{\text{old}}}(a_t | s_t)}.
\]
The clipped surrogate objective is given by:
\[
\mathcal{L}^{\pi}(\theta) = \mathbb{E}_t \left[ \min \left( r_t(\theta) \hat{A}_t, \, \text{clip}(r_t(\theta), 1 - \epsilon, 1 + \epsilon) \hat{A}_t \right) \right],
\]
where $\hat{A}_t$ is an estimator of the advantage function at time step $t$, and $\epsilon$ is a hyperparameter that controls the clipping range (typically $\epsilon = 0.1$ or $0.2$). To compute $\hat{A}_t$, PPO uses the Generalized Advantage Estimation algorithm~\cite{Schulman2016_GAE}, which relies on a learned value function $V_\phi(s)$ parameterized by $\phi$. This value function is trained to minimize the squared error between predicted and empirical returns:
\[
\mathcal{L}^{\text{V}}(\phi) = \mathbb{E}_t \left[ \left( V_\phi(s_t) - G_t \right)^2 \right],
\]
where $G_t$ is the cumulative discounted return.

\section{Dynamic Loss Weighting}\label{app:DynaWeight}

In Equation 4, we define the total loss optimized by our algorithm as the sum of the losses from PPO and the language objectives. Combining many different losses can be problematic, as different losses may take values of different orders of magnitude, which can prevent all losses from being optimized at the same rate. This is a well-known problem in the multi-task learning literature, that unfortunately does not have a widely recognized preeminent solution~\cite{Vandenhende2021_MultiTask}. Intuitively, all losses should be of the same magnitude to ensure that they contribute equally to the total loss. Knowing that all losses evolve at very different rates, and inspired from previous work~\cite{Liu2019_MTAN}, we dynamically update the weighting parameters $\beta^\pi$, $\beta^V$, $\beta^{capt}$, and $\beta^{CLIP}$ so they all have values close to 1. To do so, during training iteration $k$, we have:
\begin{equation}
    \beta^X_k=\frac{1}{L^X_{k-1}(\theta)},
\end{equation}
for each loss $L^X(\theta)$. In other words, each loss is normalized by the value of the loss at the last iteration of training. We found that it was a simple solution for weighting many very different losses, which yielded the best results in our case. Note that this strategy is also applied to the learning of grounding objectives in emergent communication baselines.

\section{Hyperparameters}

\begin{table}[h]
    \centering
    \caption{Hyperparameters used for training the different agent versions.}
    \label{app:App_LAMAC_hyperparam}
    \begin{tabular}{|cccccc}
        \hline
        \multirow{2}{*}{Hyperparameter}  & \multicolumn{5}{|c|}{Algorithm} \\ \cline{2-6} 
                        & \multicolumn{1}{|c}{No Comm}     & \multicolumn{1}{c}{EC}     & \multicolumn{1}{c}{EC-AutoEncode} & EC-LangGround & \multicolumn{1}{c|}{Lang} \\ \hline 
        
        \multicolumn{6}{c}{Language training}  \\   \hline
        \multicolumn{1}{|c|}{Lang. batch size}   & - & - & - & - & \multicolumn{1}{c|}{$1024$}  \\
        \multicolumn{1}{|c|}{Embedding dimension}   & - & - & - & - & \multicolumn{1}{c|}{$4$}  \\
        \multicolumn{1}{|c|}{Learning rate} & - & - & - & - & \multicolumn{1}{c|}{$0.007$}  \\   \hline
        
        \multicolumn{6}{c}{Agent architecture}  \\   \hline 
        \multicolumn{1}{|c|}{Context dimension $C$}       & -  & 2     & 2      & 4  & \multicolumn{1}{c|}{$16$}     \\ 
        \multicolumn{1}{|c|}{Hidden dimension $H$}   & \multicolumn{5}{c|}{$128$}\\
        \multicolumn{1}{|c|}{Nb. of hidden layers}   & \multicolumn{5}{c|}{$2$}  \\  \hline
        
        \multicolumn{6}{c}{PPO training}  \\   \hline
        \multicolumn{1}{|c|}{Nb. of parallel rollouts} & \multicolumn{5}{c|}{$250$}  \\ 
        \multicolumn{1}{|c|}{Learning rate} & \multicolumn{5}{c|}{$0.0005$}  \\ 
        \multicolumn{1}{|c|}{Nb. of PPO epochs} & \multicolumn{5}{c|}{$15$}  \\ 
        \multicolumn{1}{|c|}{Nb. of mini-batches} & \multicolumn{5}{c|}{$1$}  \\
        \multicolumn{1}{|c|}{Nb. of warm-up steps} & \multicolumn{5}{c|}{$50000$}  \\ \hline
    \end{tabular}
\end{table}

In Table~\ref{app:App_LAMAC_hyperparam}, we present the main hyperparameters used in the experiments. For training with PPO, many other hyperparameters are used, but we show only the ones that differ from the original implementation. Specific hyperparameters are defined for training the language modules: the sizes of the batches, the dimension of the embedding layers in the decoder and language encoder, and the learning rate applied on these modules. In the agent architecture, the context dimension $C$ is the dimension of the output of the communication policy. For the training of agents with the objective of PPO, we use 250 parallel rollouts: i.e., 250 parallel episodes run between each training phase. Each training phase has 15 consecutive training updates with 1 mini-batch (i.e., all the data collected during rollouts is used in one single batch). Finally, we begin training with a "warm-up" phase of 50000 steps, during which the learning rate is lowered by a factor of $10^{-2}$.

\section{Details on experimental settings}

In the MA-gym environment~\cite{Koul2019_magym}, we experiment with three distinct tasks. The first task is the classical Predator-Prey setting, showcasing a team of "predator" agents that need to find two wandering preys and coordinate to catch them. Catching a prey generates a high positive reward. Catching the two preys provokes the end of the episode. At each time step, agents receive a penalty of -1 to reward faster strategies. Agents observe their current absolute position in the grid and the colors (in RGB format) of the cells in a five-by-five area around them. In this setting, the language describes the location of observed preys, with the following vocabulary:
\begin{center}
    $V=\left\{\text{Prey, North, South, East, West, Center}\right\}.$
\end{center}

The second task is an instance of the Level-Based Foraging setting, in which agents gather resources scattered around the map. Resources have three different levels, associated with different colors, indicating the number of agents required to be foraged (from one to three agents). Higher-level resources give higher rewards. When all resources have been retrieved, the episode stops. Again, a -1 penalty is given to the agents at each time step. Observations are defined as in the Predator-Prey setting. Here, the language describes the location of observed resources, termed "Gems", with the following vocabulary:
\begin{center}
    $V=\left\{\text{Gem, North, South, East, West, Center, Purple, Green, Yellow}\right\}.$
\end{center}

The third setting is a Coordinated Placement game, where two agents have to navigate on top of landmarks (two-by-two colored areas) of the same color to win the game. The reward is defined only by the -1  penalty at each time step. Agents observe only their absolute position and the color of the cell they are in. The language allows describing the color and locations of observed landmarks, with the following vocabulary:
\begin{center}
    $V=\left\{\text{North, South, East, West, Center, Red, Green, Blue, Yellow, Cyan, Purple}\right\}.$
\end{center}

Next, we experiment in the continuous multi-agent particle environment (MPE)~\cite{Lowe2017_MADDPG}. The first task is a Predator-Prey setting with two agents and one prey. Agents are penalized at each step with the average distance between them and the prey. Agents observe their absolute position and the relative position of other entities, only if they are in observation range. Like in the MA-gym Predator-Prey setting, the language describes the positions of observed preys, with the same vocabulary.

Finally, we study an augmented version of the "simple\_reference" task in MPE, with two agents having to navigate and place themselves on top of landmarks. The goal landmark of each agent is known only to their partner, so they must communicate to solve the task. In our version, landmarks have different colors, which forces agents to communicate these colors and then search for a landmark with the right color. In each episode, three landmarks are spawned with random colors and positions. Agents are penalized at each step with the distance to their goal landmark. Agents observe their absolute position, the relative position of landmarks in their observation range, and the position and color of their partner's goal landmark. In this setting, the language describes the position and color of the observed goal landmark, with the following vocabulary:
\begin{center}
    $V=\left\{\text{North, South, East, West, Center, Red, Green, Blue}\right\}.$
\end{center}

\section{Embedding visualizations}

\begin{figure}[h]
    \centering
    \includegraphics[width=\linewidth]{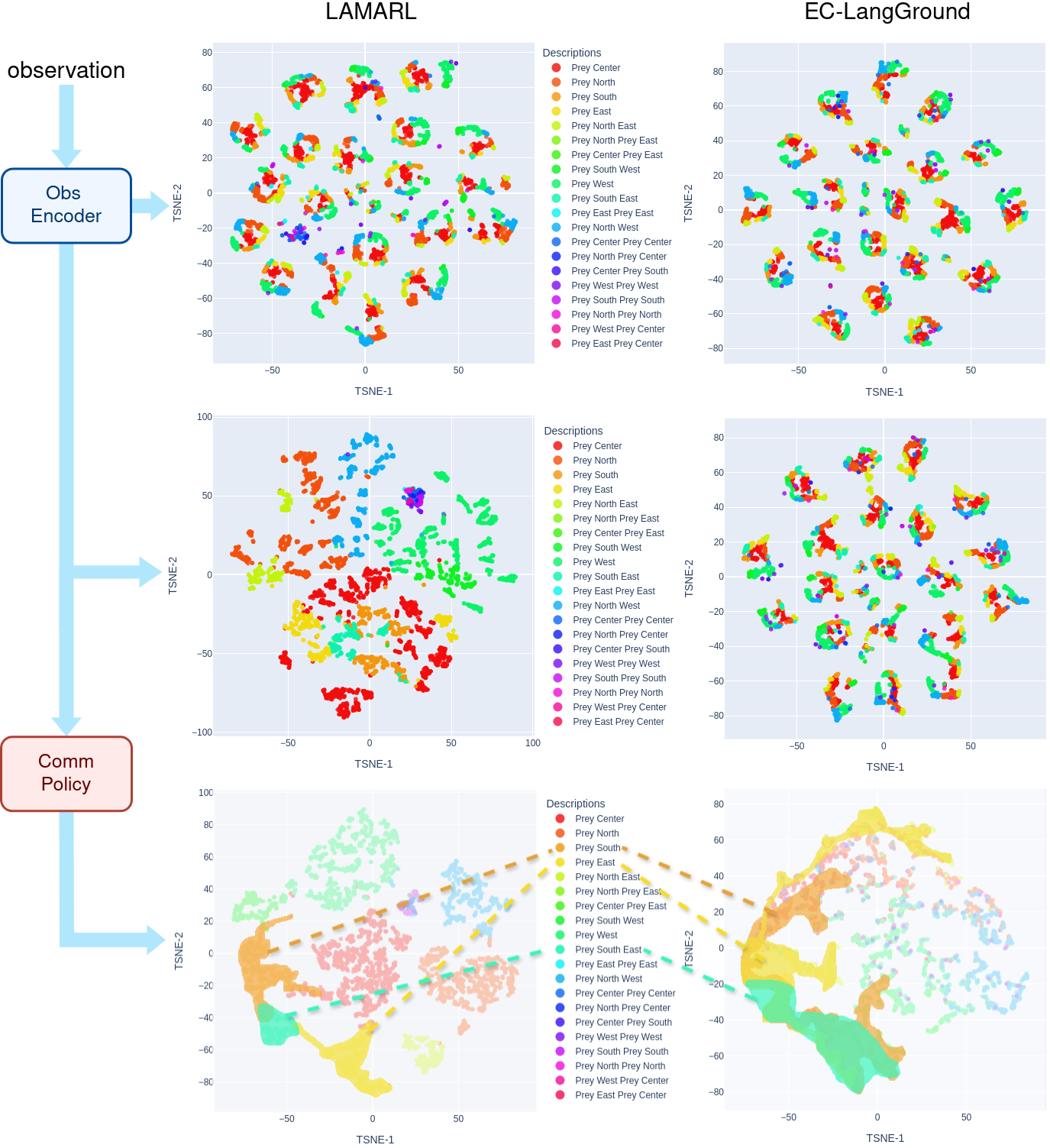}
    \caption{Visualizations of embeddings produced by subsequent layers in the agent architecture in \texttt{LAMARL} (left) and \texttt{EC-LangGround} (right). The top row shows the embedding space after the input multi-layer perceptron of the observation encoder. The middle row shows the embedding space after the observation encoder. The bottom row shows the embedding space after the communication policy (same as the figure in the main paper). To illustrate the difference in clustering quality, we highlighted the regions corresponding to descriptions "Prey South" (orange), "Prey East" (yellow), and "Prey South East" (green). }
    \label{fig:embeds_steps}
\end{figure}

Figure~\ref{fig:embeds_steps} displays the embedding spaces in subsequent layers of the agent architecture, showing the learned representations of the local observations. The first row shows the embeddings after being passed through the input multi-layer perceptron of the observation encoder. At this depth, both \texttt{LAMARL} and \texttt{EC-LangGround} have learned similar structures with cluster not primarily defined by the language labels. The second row shows the embeddings generated by the observation encoder (after a gated recurrent unit). At this level, \texttt{LAMARL} has learned to better structure its representations following the language descriptions. This demonstrates how language grounding in \texttt{LAMARL} has a deeper effect on the internal representations in the agents' networks, thus providing better guidance. Finally, the bottom row displays the embeddings generated by the communication policy, as shown in Figure 4 of the main paper. We highlight the regions of the embedding spaces that correspond to descriptions "Prey South" (orange), "Prey East" (yellow), and "Prey South East" (green), to showcase the difference between \texttt{LAMARL} and \texttt{EC-LangGround} in the quality of the learned representations. \texttt{LAMARL} features clearly defined clusters arranged coherently with regards to the spatial layout of the environment: i.e., "Prey South East" between "Prey South" and "Prey East". On the other hand, \texttt{EC-LangGround} features overlapping clusters with no clear spatial arrangement. 



\clearpage
\section{Action probabilities with messages from external sources}

\begin{figure}[h]
    \centering
     \includesvg[width=0.4\linewidth]{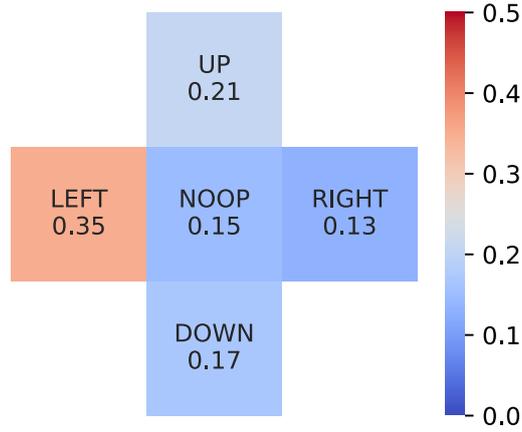}
    \caption{Action probabilities of agents evaluated in the interaction experiment, when no message is received.}
    \label{fig:Interact_nomess}
\end{figure}

\begin{figure}[h]
    \centering
     \includesvg[width=\linewidth]{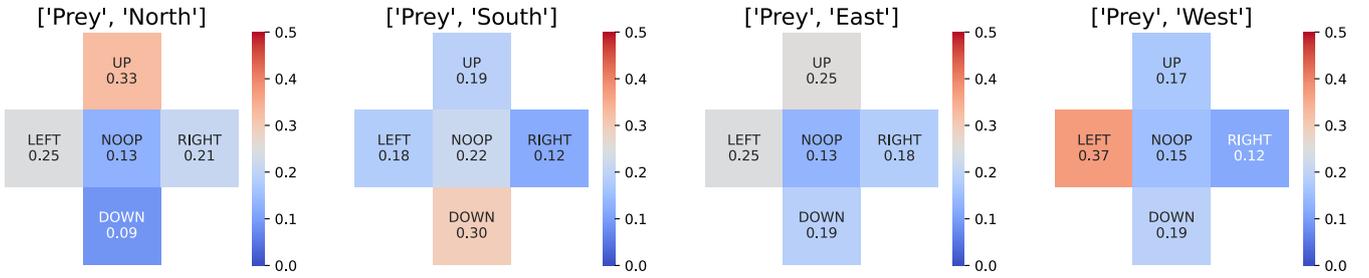}
    \caption{Action probabilities of agents evaluated in the interaction experiment, when corresponding messages are received.}
    \label{fig:Interact_mess}
\end{figure}

The changes in action probabilities displayed in Figure 7 of the main paper are computed as:
\begin{equation}
    c(a)=\frac{p^{message}(a)}{p^{no\ message}(a)} - 1, 
\end{equation}
with $p^{message}(a)$ given by Figure~\ref{fig:Interact_mess} and $p^{no\ message}(a)$ by Figure~\ref{fig:Interact_nomess}.

\end{document}